\documentclass[12pt]{article}
\usepackage{epsfig}
\usepackage{graphics}
\usepackage{axodraw}
\usepackage{graphicx}
\usepackage{amsmath}
\usepackage{amsfonts}
\usepackage{amssymb}

\newcommand{\comment}[1]{}
\def\be{\begin{equation}}
\def\ee{\end{equation}}
\def\beq{\begin{equation}}
\def\eeq{\end{equation}}
\def\bea{\begin{eqnarray}}
\def\eea{\end{eqnarray}}
\def\ba{\begin{array}{rcl}}
\def\ea{\end{array}}

\def\gv{g_V}
\def\ga{g_A}
\def\agr{{\mathrm Re} a_\gamma}
\def\azr{{\mathrm Re} \Delta a_Z}
\def\agi{{\mathrm Im} a_\gamma}
\def\azi{{\mathrm Im} \Delta a_Z}
\def\bgr{{\mathrm Re} b_\gamma}
\def\bzr{{\mathrm Re} b_Z}
\def\bgi{{\mathrm Im} b_\gamma}
\def\bzi{{\mathrm Im} b_Z}
\def\bgtr{{\mathrm Re} \tilde b_\gamma}

\def\bztr{{\mathrm Re} \tilde b_Z}

\def\bgti{{\mathrm Im} \tilde b_\gamma}

\def\bzti{{\mathrm Im} \tilde b_Z}

\def\peff{P^{\rm eff}_L}

\def\dsp{\displaystyle }

\begin{document}
\begin{flushright}
\end{flushright}

\begin{center}
\boldmath
{\Large \bf Decay-lepton correlations as probes \\[2mm] 
of anomalous $ZZH$ and $\gamma ZH$
interactions \\[4mm] in $e^+e^- \to HZ$ with polarized beams
}
\vskip 1cm
{\large Saurabh D. Rindani and Pankaj Sharma}\\
\smallskip
\smallskip
{\it Theoretical Physics Division, Physical Research Laboratory \\
Navrangpura, Ahmedabad 380 009, India}
\end{center}
\bigskip
\bigskip
\centerline{\bf Abstract}
\bigskip
\begin{quote}
We examine the contributions of various couplings in 
general $ZZH$ and $\gamma ZH$ 
interactions arising from new
physics to the Higgs production 
process $e^+e^- \to HZ$, followed by the decay of the $Z$
into a charged-lepton pair.
We take into account possible longitudinal or transverse beam
polarization likely to be available at a linear collider. 
We show how expectation values of certain simple observables
in suitable combinations with appropriate longitudinal beam
polarizations can be used to disentangle various couplings from one
another. 
Longitudinal polarization can also improve the sensitivity for measurement of
several couplings.
A striking result is that using transverse polarization, 
one of the $\gamma ZH$ couplings, 
not otherwise accessible, can be determined
independently of all other couplings.
\end{quote}
\section{Introduction}

Despite the dramatic success of the standard model (SM), 
an essential component 
of SM responsible for generating masses in the theory, viz., the Higgs 
mechanism, as yet remains untested. The SM Higgs boson, signalling
symmetry breaking in SM by means of one scalar doublet of $SU(2)$, is
yet to be discovered. A scalar boson with the properties of the SM Higgs
boson is likely to be discovered at the Large Hadron Collider
(LHC). However, there are a number of scenarios beyond the standard
model for spontaneous symmetry
breaking, and ascertaining the mass and other properties of the
scalar boson or bosons is an important task. This task would prove
extremely difficult for LHC. However, scenarios beyond SM, with 
more than just one Higgs doublet, as in the case of minimal supersymmetric
standard model (MSSM), would be more amenable to discovery at a linear $e^+e^-$
collider operating at a centre-of-mass (c.m.) energy of 500 GeV. We are at
a stage when the International
Linear Collider (ILC), seems poised to become a reality \cite{LC_SOU}. 

Scenarios going beyond the SM mechanism of symmetry breaking, and
incorporating new mechanisms of CP violation, have also become a
necessity in order to understand baryogenesis which resulted in the
present-day baryon-antibaryon asymmetry in the universe. In a theory
with an extended Higgs sector and new mechanisms of CP violation, the
physical Higgs bosons are not necessarily eigenstates of CP. 
In such a case, the production of a physical Higgs can proceed through
more than one channel, and the interference between two channels can
give rise to a CP-violating signal in the production.

Here we consider in a general model-independent way the production of a
Higgs mass eigenstate $H$ in a possible extension of SM 
through the process $e^+e^- \to HZ$ mediated
by $s$-channel virtual $\gamma$ and $Z$, followed by the decay of the
$Z$ to a final state of a charged-lepton pair, different from $e^+e^-$. This is
an important mechanism for the production of the Higgs, the other
important mechanisms being $e^+e^- \to e^+e^- H$ and $e^+e^- \to \nu
\overline \nu H$, proceeding via $e^+e^- \to HZ$ and vector-boson fusion. 
At the lowest order in SM,
$e^+e^- \to HZ$ is mediated by
$s$-channel exchange of $Z$ with a point-like
$ZZH$ vertex.
Higher-order effects or interactions 
beyond SM can modify this point-like vertex as considered in
\cite{zerwas}-\cite{cao}. 
There is also a diagram with a photon
propagator and an anomalous $\gamma ZH$ vertex.
Such  anomalous $\gamma ZH$
couplings were considered earlier in \cite{hagiwara,hikk}.
Ref. \cite{RR}  considered a four-point $e^+e^-HZ$ coupling,
which could include 
contributions of three-point $\gamma ZH$ and $ZZH$ vertices,
as well as of additional couplings going beyond $s$-channel exchanges.

Assuming Lorentz invariance, the
general structure for the vertex corresponding to the process
$V_\mu^* (k_1) \to Z_\nu (k_2) H$, where $V\equiv \gamma$ or $Z$, can
be written as \cite{hikk,skjold,biswal}
\be\label{couplings}
\Gamma^V_{\mu\nu} = g_Vm_Z \left[ a_V\,g_{\mu\nu} + \frac{b_V}{m_Z^2}\,
	(k_{1\nu}k_{2\mu} - g_{\mu\nu}k_1\cdot k_2) + 
	\frac {\tilde b_V}{m_Z^2}\,\epsilon_{\mu\nu\alpha\beta}
	k_1^\alpha k_2^\beta\right],
\ee
where $a_V$, $b_V$ and $\tilde b_V$, are form factors, which are in
general complex. The constant
$g_Z$ is chosen to be $g/\cos\theta_W$, 
so that $a_Z=1$ for SM. $g_\gamma$ is chosen to be $e$. 
Of the interactions in (\ref{couplings}), the terms with 
$\tilde b_Z$ and $\tilde b_\gamma$ are CP 
odd, whereas the others are CP even.
Henceforth we will write $a_Z=1+ \Delta a_Z$, $\Delta a_Z$ being the
deviation of $a_Z$ from its tree-level SM value. 
The other form factors are vanishing in SM at tree level. Thus the above
``couplings'', which are deviations from the tree-level SM values, could
arise from loops in SM or from new physics beyond SM. We could of course
work with a set of modified couplings where the anomalous couplings denote
deviations from the tree-level values in a specific extension of the SM
model, like a concrete two-Higgs doublet  model. The corresponding
modifications are trivial to incorporate.


In an earlier paper \cite{PS1}, we studied the sensitivity of certain angular
asymmetries of the $Z$ in $e^+e^-\to HZ$ with longitudinal and transverse
beam polarization in constraining the anomalous $\gamma ZH$ and $ZZH$
vertices. The present work is a natural extension of \cite{PS1} in terms
of including the decay of the $Z$ and considering observables which
depend on the momenta of the $Z$ decay products, which lead to new
results.

We neglect terms quadratic in anomalous couplings
assuming that the new-physics contribution is small compared to
the dominant SM contribution.
We include the possibility that the beams have polarization, either
longitudinal or transverse.

We are thus addressing the question of how well the form factors for the
anomalous $ZZH$ and $\gamma ZH$ 
couplings in $e^+e^-\to HZ$ can be {\em simultaneously} 
determined from the measurement of simple
observables constructed out of the momenta of the $e^{\pm}$ and of the
charged-lepton pair arising in the decay of the $Z$
utilizing unpolarized beams and/or polarized beams.
This question,
taking into account a new-physics contribution which merely modifies the
form of the 
$ZZH$ vertex, has been addressed before in several works
\cite{zerwas,skjold,biswal,cao}. 
This amounts to assuming that the $\gamma ZH$ couplings are zero or
negligible. Refs. \cite{hagiwara,hikk} do take into account both
$\gamma ZH$ and $ZZH$ couplings. However, they relate both
to coefficients of terms of higher dimensions in an effective
Lagrangian, whereas we treat all couplings as independent of one
another. Moreover, Gounaris et al. \cite{hagiwara} do 
not discuss effects of beam
polarization. On the other hand, we attempt to seek ways to determine the
couplings completely independent of one another.
Refs. \cite{hikk} do have a similar approach to ours. They make use of
optimal observables \cite{optimal}  and consider only longitudinal electron
polarization,
whereas we seek to use simpler observables
and consider the effects of longitudinal and transverse
polarization of both $e^-$ and $e^+$ beams. The first paper of \cite{hikk}
also includes $\tau$ polarization and $b$-jet charge
identification which we do not require. 

One specific practical
aspect in which our approach differs from
that of the effective Lagrangians is that while the couplings are all
taken to be real in the latter approach, we allow the couplings to be
complex, and in principle, momentum-dependent form factors.

Polarized beams are likely to be available at a linear collider, and
several studies have shown the importance of linear
polarization in reducing backgrounds and improving the sensitivity to
new effects \cite{gudi}. The question of whether transverse beam
polarization, which could be obtained with the use of spin rotators,
would be useful in probing new physics, has been addressed in recent
times in the context of the ILC 
(see \cite{gudi},\cite{basdrtt} and references therein).


When all couplings are assumed to be
independent and nonzero, our observables are linear
combinations of a certain number of anomalous couplings 
(in our approximation of neglecting
terms quadratic in anomalous couplings). By using that number of
expectation values, for example, of different observables, or of 
the same observable
measured for different beam polarizations, one can solve simultaneous
linear equations to determine the couplings involved.
This is the approach we wish to emphasize here.
A similar technique of considering combinations of different
polarizations was made use of, for example, in \cite{poulose}.

We find that longitudinal polarization is useful in
giving information on a different combination of
couplings as compared to the unpolarized case, thus allowing a simultaneous
determination of couplings using polarized and unpolarized data. 
Transverse polarization is found useful in isolating contributions
of different couplings to certain observables. One specific coupling,
viz., $\agi$, can only be determined
with the use of transverse polarization, and a particular observable is
found to enable its measurement independent of all other couplings.
Unpolarized or longitudinally
polarized beams provide no access to $\agi$.
Transverse polarization can also help isolate other couplings, since, as
it turns out, usually the contribution of one coupling dominates most
observables.
 
\comment{
It may be appropriate to contrast our approach with the usual effective
Lagrangian approach. In the latter approach, it is
assumed that SM is an effective theory which is valid up to a cut-off scale
$\Lambda$. The new physics occurring above the scale of the cut-off may
be parameterized by higher-dimensional operators, appearing with powers
of $\Lambda$ in the denominator. These when added to the SM Lagrangian
give an effective low-energy Lagrangian
where, depending on the scale of the momenta involved, one
includes a range of higher-dimensional operators up to a certain maximum
dimension. Our effective theory is not a low-energy limit, so that the
form factors we use are functions of momentum not restricted to low
powers. Thus, we have introduced $m_Z$  rather than a cut-off scale, 
just to make the form factors dimensionless. 
}

\def\plm{p_{l^-}}
\def\plp{p_{l^+}}
\def\dsp{\displaystyle}
\def\gzh{$\gamma ZH$~}
\def\zzh{$ZZH$~}

In the next section we discuss how model-independent
$ZZH$ and $\gamma ZH$ couplings contribute to the process $e^+e^- \to
HZ$ with polarized beams. 
Section 3 deals with observables whose
expectation values can be used
for separating various form factors and Section 4 describes the numerical
results.  Section 5
contains our conclusions and a discussion.

\section{\boldmath Contribution of anomalous couplings to 
the process $e^+e^- \to HZ$}

We consider the process 
\begin{equation}\label{process}
e^- (p_1) + e^+ (p_2) \to Z^\alpha (q) + H(k) \to \ell^+ (p_{l^+})+
\ell^- (p_{l^-})+ H(k),
\end{equation}
where $\ell$ is either $\mu$ or $\tau$. 
Helicity amplitudes for the process were obtained earlier in the context
of an effective Lagrangian approach \cite{hagiwara,hikk}.
We have used instead trace techniques employing the
symbolic manipulation program `FORM' \cite{form}.
We neglect the mass of the electron. 

We choose the $z$ axis to be the direction of the $e^-$ momentum, and
the $xz$ plane to coincide with the plane of the momenta of $e^-$ and $\ell^-$
in the case
when the initial beams are unpolarized or longitudinally polarized. 
The positive $x$
axis is chosen, in the case of transverse polarization, to be along the
direction of the $e^-$ polarization, and the $e^+$ polarization is taken
to be parallel to the $e^-$ polarization.

The details of the analytical expressions for the differential cross
sections in the presence of longitudinal and transverse polarizations
will be given elsewhere. Here, we list merely salient features of the
results.
 
The differential cross section with longitudinally polarized beams,
apart from  an overall factor $(1-P_L\overline P_L)$, depends on the
``effective polarization'' $\peff
= \frac{P_L - \overline P_L}{1-P_L\overline P_L}$,
where $P_L$, $\overline P_L$ are the longitudinal polarizations of the
electron and positron beams, respectively.
Since $\peff$ is about 0.946 for $P_L=0.8$, $\overline P_L=-0.6$, and 0.385
for $P_L=0.8$, $\overline P_L=0.6$, a high degree of effective
polarization can be achieved using these partial polarizations for $e^-$ and
$e^+$ beams opposite in sign to each other, which are expected to be
available at the ILC.

Including the decay of $Z$ into charged leptons gives us additional 
contributions 
 from anomalous couplings $\bgi$, $\bzi$, $\bgtr$ and $\bztr$, which are absent in 
 the distributions without $Z$ decay considered in \cite{PS1}.
Couplings $\agi$ and $\azi$ are still absent from distributions of Z for longitudinally 
polarized beams even though we include $Z$ decay.

The expression for the differential cross section with transverse polarization 
$P_T$ for $e^-$ beam 
and $\overline{P}_T$ for $e^+$ beam has terms either independent of the
$P_T$ and $\overline P_T$, or proportional to the product $P_T\overline
P_T$. With transverse polarization, we do get a contribution from
$\agi$, which is missing with unpolarized or longitudinally polarized
beams.

\section{Observables}
We have evaluated the expectation values of observables $X_i$
$(i=1,2,\ldots 8)$ for unpolarized and
longitudinally polarized beams, and observables $Y_i$ $(i=1,2,\ldots 6)$
for transversely polarized beams. 
$X_1$-$X_8$ 
are sensitive 
to longitudinal beam polarization and $Y_1$-$Y_6$ to transverse beam 
polarization. 
The definitions of the 
observables $X_i$ and $Y_i$ are
found respectively in Tables \ref{longlim500} and \ref{translim500}.

An obvious choice of observable is the total cross section $\sigma$,
which is even under C, P and T\footnote{Henceforth, T will always refer
to {\em naive} time reversal, i.e., reversal of all momenta and spins,
without interchange of initial and final states.}. In the
presence of anomalous couplings, this gets contribution from $\azr,
\agr, \bzr$ and $\bgr$. 
The cross section with longitudinally polarized beams can lead to 
stringent limits on the couplings. However, since the results are
subject to higher-order corrections \cite{comelli}, 
which we do not include, we will
concentrate on expectation values of observables. These being ratios,
will be less sensitive to higher-order corrections.

Observables which are even under 
CP get contributions from $\Delta a_Z$, $a_{\gamma}$, $b_Z$ 
and $b_{\gamma}$, and observables which are odd under CP get
contributions from  
$\tilde{b}_Z$ and $\tilde{b}_{\gamma}$. The CPT theorem implies 
that observables which are CP even and T even or CP odd and T odd 
would get contribution from real parts of couplings.
On the other hand, observables which are 
CP odd and T even or CP even and T odd get 
contributions from imaginary part of the couplings.

The observables we have chosen are by no means exhaustive. They  have
been chosen based on simplicity, and with the idea of extracting
information on all couplings, and if possible, placing limits on
them independently of one another.

\section{Numerical Calculations}

For the purpose of numerical calculations, we have made use of the following values of parameters: 
$M_Z=91.19$ GeV, $\alpha(M_Z)=1/128$, $\sin^2\theta_W=0.22$. 
We have evaluated expectation values of 
the observables and their sensitivities to the various anomalous couplings 
for a linear collider operating 
at $\sqrt{s}=500$ GeV having integrated luminosity 
$\int\mathcal Ldt=500$ fb$^{-1}$. We have assumed 
longitudinal polarizations of $P_L=\pm 0.8$ and $\overline{P}_L=\pm 0.6$ 
would be accessible for $e^-$ 
and $e^+$ beams respectively, and identical degrees of transverse
polarization.

 
We have examined the accuracy to which couplings 
can be determined 
from a measurement of the correlations of observables $O_i$.
The limits which can be placed at the $95\%$ CL on a coupling 
contributing to the correlation of  
$O_i$ is obtained from
\beq\label{lim}
|\langle O_i\rangle-\langle O_i\rangle_{\rm SM}|=f\,
\frac{\sqrt{\langle O_i^2\rangle_{\rm SM}-
\langle O_i\rangle^2_{\rm SM}}}{\sqrt{L\sigma_{\rm SM}}},
\eeq 
where the subscript ``SM" refers to the value in SM,
and where $f$ is 1.96 when only one coupling is
assumed non-zero, and 2.45 when two couplings contribute.

Since polarized beams would not be available for the full
period  of operation of the collider,
we consider alternative options of luminosities for which individual
combinations of polarization would be used. We 
consider that the collider would be run in three phases with 
three different combinations 
$(P_L,\overline{P}_L)$
of beam 
polarizations with values 
$(0,0)$, $(+0.8,-0.6)$ and $(-0.8,+0.6)$, respectively with  integrated
luminosities of 
$250$ fb$^{-1}$, $125$ fb$^{-1}$, and $125$ fb$^{-1}$.

Let us discuss how limits may be obtained using each observable and with 
various combinations of 
beam polarizations in some detail.

\subsection{Sensitivities with unpolarized beams}

Each observable $X_i$ chosen by  us has dependence
on a combination of a limited number of couplings, dependent on CP and T
properties.
Thus a single observable
can only be used to determine, or put limits on, a combination of
couplings. We can determine, from a single observable, 
limits on individual couplings either under
an assumption on the remaining couplings which contribute to the
observable, or by combining the results from more than one observable,
or from more than one combination of polarization.
We will refer to limits on a coupling as an individual limit if the
limit is obtained on the assumption of all other couplings being zero.
If no such assumption is made, and more than one observable is used
simultaneously to put limits on all couplings contributing to these
observables, we will refer to the limits as simultaneous limits.
\begin{table}[t]
\centering
\begin{tabular}{cccccc}
\hline
\hline
\multicolumn{3}{c}{} & \multicolumn{3}{c}{Limits for polarizations}
\\

\multicolumn{2}{c}{Observable} & Coupling  & $P_L=0$ & $P_L=0.8$ & $P_L=0.8$\\ 
       &            &            & $\overline{P}_L=0$ &
$\overline{P}_L=0.6$ & $\overline{P}_L=-0.6$\\ 
\hline
 $X_1$  &$(p_1-p_2).q$           
  & $\bzti $    & $4.11\times 10^{-2} $ & $8.69\times 10^{-2} $&
$9.94\times 10^{-3}$\\
  & & $\bgti $  & $1.49\times 10^{-2} $ &$2.06\times 10^{-2} $
&$1.22\times 10^{-2} $ \\

 $X_2$       & $P.(p_{l^-}-p_{l^+})$         
 & $\bzti $     & $4.12\times 10^{-2} $ & $5.99\times 10^{-2} $ &
$3.84\times 10^{-2} $\\
 & & $\bgti $   & $5.23\times 10^{-1} $ & $3.12\times 10^{-1} $&
$5.52\times 10^{-2} $\\

 $X_3$  & $(\vec{p}_{l^-}\times \vec{p}_{l^+})_z$ 
 & $\bztr $           & $1.41\times 10^{-1} $ & $2.97\times 10^{-1} $ &
$3.40\times 10^{-2} $\\
 & & $\bgtr $ & $5.09\times 10^{-2} $ & $7.05\times 10^{-2} $ &
$4.15\times 10^{-2} $ \\

 $X_4$  & $(p_1-p_2).(p_{l^-}-p_{l^+})$
 & $\bztr $     & $2.95\times 10^{-2} $ & $4.29\times 10^{-2} $ &
$2.75\times 10^{-2} $\\
& $\times (\vec{p}_{l^-}\times \vec{p}_{l^+})_z$      
  & $\bgtr $   & $3.81\times 10^{-1} $ &$ 2.24\times 10^{-1}$ &
$3.95\times 10^{-2}$ \\

 $X_5$  & $(p_1-p_2).q(\vec{p}_{l^-}\times
\vec{p}_{l^+})_z$      
 & $\bzi $      & $ 7.12\times 10^{-2}$ & $ 1.04\times 10^{-1}$ & $
6.64\times 10^{-2}$\\
 & & $\bgi $    & $ 9.10\times 10^{-1}$ & $ 5.42\times 10^{-1}$ &
$9.53\times 10^{-2}$ \\

 $X_6$       & $P.(p_{l^-}-p_{l^+})(\vec{p}_{l^-}\times \vec{p}_{l^+})_z$
 & $\bzi $      & $ 7.12\times 10^{-2}$ & $ 1.50\times 10^{-1}$ & $
1.72\times 10^{-2}$\\
 &       
& $\bgi $    & $ 2.58\times 10^{-2}$ & $ 3.57\times 10^{-2}$ &
$2.10\times 10^{-2}$ \\

 $X_7$       & $[(p_1-p_2).q]^2$  
 & $\bzr $      & $1.75\times 10^{-2} $ & $2.54\times 10^{-2} $&
$1.63\times 10^{-2} $\\
 &
 & $\bgr $    & $2.23\times 10^{-1} $ &$ 1.34\times 10^{-1}$ &
$2.35\times 10^{-2}$ \\ 

 $X_8$       & $[(p_1-p_2).(\plm-\plp)]^2$   
 & $\bzr $      & $1.53\times 10^{-2} $ & $2.22\times 10^{-2} $&
$1.42\times 10^{-2} $\\
 & 
& $\bgr $    & $1.94\times 10^{-1} $ &$ 1.16\times 10^{-1}$ &
$2.04\times 10^{-2}$ \\ 

\hline
\hline
\end{tabular} 
\caption{\label{longlim500} The $95$ \% C.L. limits on the anomalous
$ZZH$ and $\gamma ZH$ couplings, 
chosen nonzero one at a time, from various observables with unpolarized and
longitudinally polarized beams 
}
\end{table}
The individual limits obtained from various observables for unpolarized
and longitudinally polarized beams are given in Table \ref{longlim500}.

We now proceed to examine in detail observables or combinations of
observables and the coupling or couplings about which they can give
information.

\begin{figure}[ht]
\centering
 \includegraphics[scale=0.65]{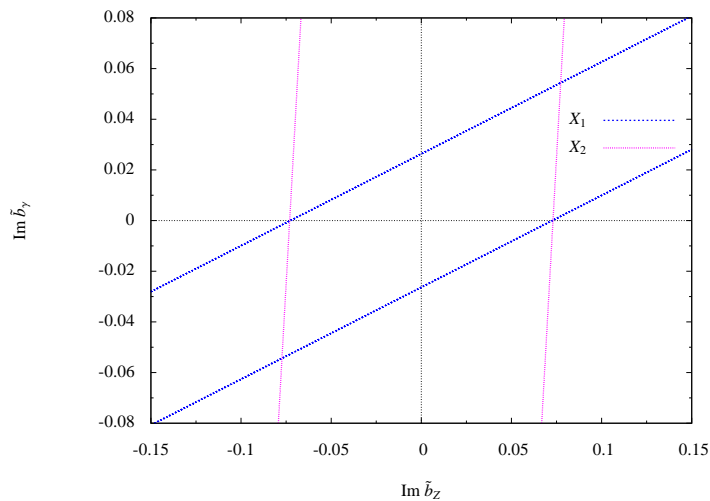}
\caption{\label{PlotO1O10}The region in the $\bzti$-$\bgti$ plane accessible at the $95$\% CL with 
observables $X_1$ and $X_2$ with unpolarized beams for integrated luminosity $L=250$ fb$^{-1}$.}
\end{figure}

$X_1$ and $X_2$ probe different combinations of $\bzti$ and
$\bgti$. 
We show in Fig.\ref{PlotO1O10}, which is sample figure, 
a plot of relation eq. (\ref{lim}) in the space of the couplings involved 
for observables $X_1$ and $X_2$ utilizing only unpolarized beams. The
intercepts on the two axes of each line give us the individual limits on
the two couplings for that observable. 
The lines 
corresponding to two combinations gives a closed region which is the 
allowed region at $95\%$ CL.
The simultaneous limits obtained by considering the extremities of 
this closed region are
\beq
|\bzti|\leq 7.73\times 10^{-2},\; |\bgti|\leq 5.44\times 10^{-2}.
\eeq
For $X_3$ and $X_4$ 
we can draw a figure analogous to Fig.\ref{PlotO1O10} 
these couplings which would lead to the simultaneous limits
\beq
|\bztr|\leq 6.08\times 10^{-2}, |\bgtr|\leq 1.12\times 10^{-1}
\eeq
Similarly, the 
simultaneous limits obtained 
from the observables $X_5$ and $X_6$ are 
\beq
|\bzi|\leq 1.25\times 10^{-1},\; |\bgi|\leq 9.39\times 10^{-2}
\eeq
Simultaneous limits on $\bzr$ and $\bgr$ 
from $X_7$ and $X_8$ are large since slopes of 
two lines corresponding to $X_7$ and $X_8$ are of same sign and approximately 
equal in magnitude.

$\azi$, $\agi$ do not appear in the differential cross section. $\azr$
and $\agr$ too cannot be determined from $X_7$ and $X_8$. This is because 
in the determination of $\langle X_{7,8} \rangle \equiv \int X_{7,8} d\sigma /
\sigma$, the contribution of $\azr$ to the numerator  
is cancelled exactly by its contribution at the linear order 
to the denominator. A similar cancellation takes place for $\agr$ approximately.  

\subsection{Sensitivities with longitudinal beam polarization}

We now consider measurement of
correlations with different combinations of longitudinal polarization. 
Since these
would give different combinations of couplings, their measurements may
be used to put simultaneous limits on couplings, without assuming any
coupling to be zero. 

\begin{figure}[h]
\centering
 \includegraphics[scale=0.65]{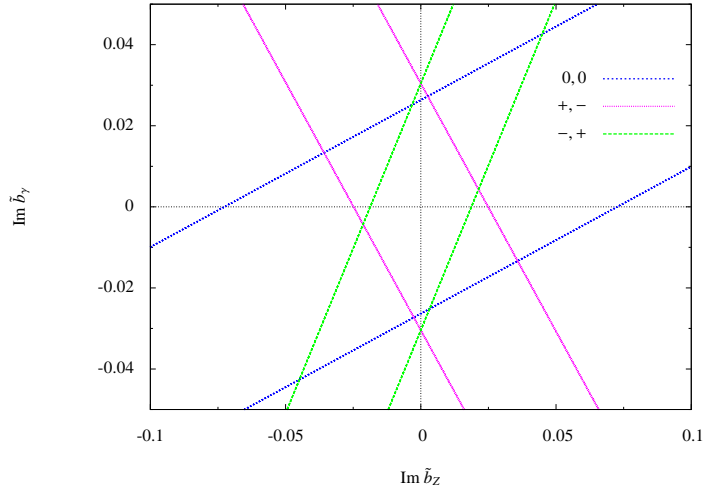}
\caption{\label{PlotO1}The region in the $\bzti$-$\bgti$ plane accessible at the $95$\% CL with observable 
$X_1$ with different beam polarization configurations. $(0,0)$, $(+,-)$ and $(-,+)$ stand for 
$(P_L,\overline{P}_L)=$ $(0,0)$, $(0.8,-0.6)$ and $(-0.8,0.6)$ respectively.}
\end{figure}

A graphical way of obtaining simultaneous limits with different
combinations of polarization is illustrated for $X_1$ in 
 Fig.\ref{PlotO1} where relation eq. (\ref{lim}) is plotted 
in the space of the
couplings involved for
unpolarized beams denoted by $(0,0)$, 
and for the two combinations of longitudinal
polarizations $(P_L,\overline{P}_L)\equiv(0.8,-0.6)$,
and $(P_L,\overline{P}_L)\equiv(-0.8,0.6)$, 
respectively denoted by $(+,-)$ 
and $(-,+)$. The lines
corresponding to any two combinations gives a closed region which is the
allowed region at $95\%$ CL.
In principle, allowed regions with other combinations of polarization
can also be plotted, and the smallest region would correspond to the
best limits.
We note that certain observables get contribution from combinations like
$\gv - \ga\peff$ in case of certain $\gamma ZH$ couplings or $2\gv\ga - (\gv^2
+ \ga^2)\peff$ in case of certain $ZZH$ couplings, where $\gv,\ga$ are
the vector and axial-vector couplings of the $Z$ to charged leptons. 
In these cases, the
sensitivity of those couplings to longitudinal polarization is high for
the reason that the polarization dependent term gets an enhancement factor
of $(\gv^2+\ga^2)/(2\gv\ga)\approx -4.2$, or $\ga/\gv \approx 8.3$ as 
compared to the unpolarized term, for the cases of $ZZH$ and $\gamma ZH$
couplings, respectively.
This enhancement occurs for 
couplings $\bzti$ for $X_1$, 
$\bgti$ for $X_2$, 
$\bztr$ for $X_3$, 
$\bgtr$ for $X_4$, 
$\bgi$ for $X_5$, 
$\bzi$ for $X_6$, and $\bgr$ for $X_7$ and 
$X_8$. 

We list in Table \ref{simlonglim} the simultaneous limits which can be
obtained using different combinations of polarizations for the various
observables. 
\begin{table}[t]
\centering
\begin{tabular}{ccccccc}
\hline\hline
& &\multicolumn{3}{c}{Limit on coupling for the} \\
Observable & Coupling &\multicolumn{3}{c}{polarization combination}\\
           &          & $(0,0),(-,+)$&$(0,0),(+,-)$&$(-,+),(+,-)$\\
\hline
$X_1$ & $\bzti$  &$4.50\times10^{-2}$ &$3.59\times10^{-2}$ &$2.14\times
10^{-2}$ \\
      & $\bgti$  &$4.28\times10^{-2}$ &$2.74\times10^{-2}$
&$3.04\times10^{-2}$ \\ 

$X_2$ & $\bzti$  &$9.73\times10^{-2}$ &$7.56\times10^{-2}$ &$8.54\times
10^{-2}$ \\
      & $\bgti$  &$3.06\times10^{-1}$ &$2.19\times10^{-1}$
&$1.37\times10^{-1}$ \\ 

$X_3$ & $\bztr$ & $1.54\times10^{-1}$ & $1.22\times10^{-1}$ & $
7.29\times 10^{-2}$ \\
	& $\bgtr$ & $1.46\times10^{-1}$ & $9.31\times10^{-2}$ &
$1.08\times10^{-1}$ \\

$X_4$ & $\bztr$ & $5.37\times10^{-2}$ & $6.89\times10^{-2}$& $6.10 \times
10^{-2}$ \\
	& $\bgtr$ & $1.56\times10^{-1}$ & $2.18\times10^{-1}$ &
$9.78\times10^{-2}$ \\

$X_5$ & $\bzi$ & $1.67\times10^{-1}$ & $1.29\times10^{-1}$& $1.48 \times
10^{-1}$ \\
	& $\bgi$ & $5.27\times10^{-1}$ & $3.76\times10^{-1}$ &
$2.36\times10^{-1}$ \\

$X_6$ & $\bzi$ & $7.79\times10^{-2}$ & $6.18\times10^{-2}$ & $3.69 \times
10^{-2}$\\
      & $\bgi$ & $7.39\times10^{-2}$ & $4.72\times10^{-2}$ & $5.27\times
10^{-2}$\\

$X_7$ & $\bzr$ & $2.53\times10^{-2}$ & $1.27\times10^{-2}$ & $3.11 \times
10^{-2}$  \\
	& $\bgr$ & $ 1.05\times10^{-1}$ & $5.74\times10^{-2}$ & $5.11\times
10^{-2}$ \\

$X_8$ & $\bzr$ & $2.58\times10^{-2}$  & $2.05\times10^{-2}$ & $3.37 \times
10^{-2}$ \\
	& $\bgr$ & $1.15\times10^{-1}$ & $6.33\times10^{-2}$ & $5.26\times
10^{-2}$ \\

\hline\hline
\end{tabular}
\caption{Simultaneous $95$ \% C.L. limits on the anomalous
$ZZH$ and $\gamma ZH$ couplings from various observables using
longitudinally polarized beams with different polarization combinations
$(0,0)$, i.e., $ P_L=0,\overline P_L=0$, $(\pm,\mp)$, i.e., $(P_L=\pm
0.8,\overline P_L=\mp 0.6)$
for $\sqrt{s}=500$ GeV and integrated luminosity $\int \mathcal L
~dt=500$ fb$^{-1}$. }\label{simlonglim}
\end{table}
%
%
\subsection{Sensitivities with transverse beam polarization}

As observed earlier, observables $Y_i$ 
have vanishing expectation values in
the absence of transverse polarization.
We now
discuss the effect of transverse polarization on these observables, 
which did not figure in the earlier discussion. 
Terms in the differential cross section dependent on transverse
polarization have the combination $P_T\overline P_T$, and hence
both beams need to have non-zero polarization to observe the effects of these
terms. 

We have listed in Table \ref{translim500} the results for individual
limits obtained following the procedures followed earlier.The most
significant result is for the coupling $\agi$. We find
that the observable $Y_1$ can constrain $\agi$
independent of all other couplings. This is particularly significant
because $\agi$ cannot be constrained with longitudinal polarization.
In the determination of $\langle Y_2 \rangle $
the numerator receives contribution only from $\azr$ and
$\agr$. However, the denominator at the linear order cancels the contribution of
$\azr$ exactly and that of $\agr$ approximately, while introducing a
dependence of $\langle Y_2 \rangle$ on $\bzr$ and $\bgr$. 
\begin{table}[t]
\begin{center}
\begin{tabular}{cccc}
\hline
\hline\\
\multicolumn{3}{c}{} & Limits for polarizations
\\

\multicolumn{2}{c}{Observable} & Coupling   & $P_T=0.8$,\,  $\overline
P_T=\pm 0.6$\\ 
\hline\\
  $Y_1$         & $(q_x q_y)$   
 & $\agi $              & $1.98\times 10^{-1}
$\\

 $Y_2$  & $(q_x^2-q_y^2)$       
 & $ \agr $       & $8.15\times 10^{-1} $\\
 && $ \bzr $       & $2.65\times 10^{-2} $\\
 && $ \bgr $       & $3.41\times 10^{-1} $\\

 $Y_3$  & $(\plm-\plp)_x(\plm-\plp)_y$  
 & $\agi $                      & $9.62 $\\
 & & $\bgi $                    & $4.72\times
10^{-2}$ \\

 $Y_5$         & $(\plm-\plp)_x(\plm-\plp)_y q_z$      
 & $\bztr $                     & $5.56\times
10^{-2} $\\
 & & $\bgtr $                   & $6.89\times
10^{-1}$ \\

 $Y_4$  & $q_xq_y(p_{l^-}-p_{l^+})_z$   
 & $\mathrm Im b_Z $             & $1.58\times 10^{-1}$\\
 & & $\mathrm Im b_{\gamma} $     & $1.96 $ \\
$Y_6$&$ [(p_{l^-})_x^2-(p_{l^+})_x^2] -[(p_{l^-})_y^2-(p_{l^+})_y^2]$
 & $\bzti$                               & $1.10\times10^{-1}$\\
& 
& $\bgti$ 				 & $1.36 $\\

 \hline
\hline
\end{tabular} 
\caption{\label{translim500}The $95$ \% C.L. limits on the anomalous
$ZZH$ and $\gamma ZH$ couplings, 
chosen nonzero one at a time from observables with 
transversely polarized beams 
for $\sqrt{s}=500$ GeV and $\int \mathcal L ~dt=500$ fb$^{-1}$. }
\end{center}
\end{table}

The other observables listed in Table \ref{translim500} do not allow limits
on single couplings to be isolated. 
However, in each of these cases, if one assumes the couplings 
contributing to the expectation value to
be of the same order of magnitude, then one of the couplings make a 
dominant contribution to the 
expectation value, leading to an independent limit on that
coupling. For example,
$Y_2$, $Y_3$, $Y_4$, $Y_5$ and $Y_6$ can place independent limits
on $\bzr$, $\bgi$, $\bztr$, $\bzi$ and $\bzti$, respectively.  


\subsection{Effect of cuts and change in centre-of-mass energy}

Since the anomalous couplings $b_Z$, $b_{\gamma}$, $\tilde b_Z$ and
$\tilde b_{\gamma}$ corresponding to interactions which are momentum
dependent, it is expected that a change in the c.m. energy would bring
about a change in the sensitivity. To investigate this possibility, 
we have obtained sensitivities of all the observables to the anomalous
couplings
 at two other center of mass 
energies i.e., $\sqrt{s}=800$ GeV with integrated luminosity $\int
\mathcal L~ dt=500$ fb$^{-1}$
and $\sqrt{s}=1000$ GeV
with integrated luminosity 
$\int \mathcal L~ dt=1000$ fb$^{-1}$.

We find that $X_1$, $X_2$ and $Y_6$ 
become less 
sensitive to couplings $\bzti$ and $\bgti$ as the c.m. energy
increases. However, 
we have better limits at $\sqrt{s}=1000$ GeV than at $\sqrt{s}=800$ GeV,
because the reduced sensitivity is compensated for by higher 
luminosity at $\sqrt{s}=1000$ GeV.
$X_3$, $X_4$, $X_5$ and $Y_4$
are more sensitive to anomalous couplings at higher
energies.
$X_7$ becomes less sensitive to couplings $\bzr$ and $\bgr$ while 
$X_8$ gives better limits to these couplings at higher energies.
Observables $Y_1$, $Y_2$ and $Y_5$ 
become less sensitive to anomalous couplings at higher
energies.
$Y_3$ behaves differently relative to all other
observables. While the limit on $\agi$ 
improves by about an order of magnitude, the limit on $\bgi$ get worse with 
 increase in c.m.  energy.

In practice, any measurement will need kinematical
cuts for the identification of the decay leptons. 
We have examined the effect on our results of the following kinematical
cuts \cite{biswal}:
\smallskip

\noindent 1. $E_f\geq10$ GeV for each outgoing charged lepton,\\
\noindent 2. $5^{\circ}\leq \theta_0\leq 175^{\circ}$ for each outgoing
charged lepton to remain away from the beam pipe,\\
\noindent 3. $\Delta R_{ll}\geq0.2$ for the pair of charged lepton, where 
       $(\Delta R)^2\equiv(\Delta \phi)^2+(\Delta \eta)^2$, $\Delta
\phi$ and $\Delta \eta$ being
the separation in azimuthal angle and rapidity,
respectively, for detection of the two leptons as separated.

 In addition to this, we impose a cut 
$|m_{l^-l^+}-M_Z|\leq 5\Gamma_Z$
on the invariant mass $m_{l^-l^+}$ of the lepton pair, so as to constrain 
 the $Z$ - boson to be more or less on shell. 
This cut would allow us to test how well our results would simulate the
results for a genuinely on-shell $Z$. Moreover, the cut would also
reduce contamination from $\gamma\gamma H$ couplings, which contribute
in principle to the process (\ref{process}), though not to $e^+e^- \to
HZ$.

After imposing these cuts, we find that all observables except
$X_1$, $X_2$ and $Y_6$ are not very 
sensitive to these cuts. The limit on  
$X_1$, $X_2$ and $Y_6$ change by $20-30$\%.

\section{Conclusions and discussion}

We have obtained angular distributions for the process $e^+e^- \to ZH 
\to  \ell^+\ell^-  H$
in the presence of anomalous $\gamma ZH$ and $ZZH$ couplings to linear
order in these couplings in the presence of longitudinal and transverse
beam polarizations. We have then looked at observables 
which can be used in combinations to disentangle the various couplings
to the extent possible. 
We have also obtained the sensitivities of these observables and
asymmetries to the various couplings for a definite configuration of the
linear collider.

In certain cases where the contribution of a
coupling is suppressed due to the fact that the vector coupling of the
$Z$ to $e^+e^-$ is numerically small, longitudinal polarization helps to
enhance the contribution of this coupling. As a result, longitudinal
polarization improves the sensitivity.
The main advantage of transverse polarization is that 
it enables 
constraining $\agi$ which is not accessible without transverse polarization. 
Moreover, it helps to
determine certain couplings independent of all other couplings.

We find that with a linear collider operating at a c.m. energy of 500
GeV with the capability of 80\% electron polarization and 60\% positron
polarization with an integrated luminosity of 500 fb$^{-1}$, 
with the observables described above, it would be
possible to place 95\% CL individual limits of the order of a few times 
$10^{-2}$ on all couplings taken 
nonzero one at a time with use of an appropriate combination ($P_L$ and
$\overline P_L$ of opposite signs) of longitudinal beam polarizations.
This is an improvement by a factor of 5 to 10 as compared to the
unpolarized case. 
The simultaneous limits possible are, as expected, less stringent. While they
continue to be better than $5\times10^{-2}$ for most couplings, they
range between $5\times10^{-2}$ and about $10^{-1}$ for $\bgr$, 
$\bztr$ and $\bgtr$.
Transverse polarization enables
the determination of $\agi$ independent of all other couplings, with a
possible 95\% CL limit of about 0.2. 
Independent limits on $\bzr$ and $\bgi$ of a few times $10^{-2}$ are
possible, whereas those on $\bzi$, $\bztr$ and $\bzti$ would be somewhat
larger, ranging upto about 0.1.    
Our procedure does not permit any limit on $\azr$ or
$\azi$, and only a weak limit on $\agr$.

We have assumed that only one leptonic decay mode of $Z$ is observed.
Including both $\mu^+\mu^-$ and $\tau^+\tau^-$ modes would trivially
improve the sensitivity. In case of observables like $X_1$, $Y_1$, $Y_2$,
which do not need charge identification, even hadronic decay modes of
$Z$ can be included, which would considerably enhance the sensitivity.

In fact, in our earlier work \cite{PS1}, where we had 
not included $Z$ decay, the
sensitivities we obtained were better simply because we did not restrict
to one decay channel. On the other hand, considering a specific
charged-lepton channel has enabled us to get a handle on  
$\bgi$, $\bzi$, $\bgtr$ and $\bztr$, which were not accessible in
\cite{PS1}.

It is appropriate to compare our results with those in works using the
same parameterization as ours for the anomalous coupling and with an
approach similar to ours. The paper of 
Han and Jiang \cite{skjold} deals with CP-violating $ZZH$
couplings,
and it is possible to compare the 95\% CL limits on $\bzti$
with those obtained by them using the
forward-backward asymmetry of the $Z$.
With identical values of $\sqrt{s}$ and integrated
luminosity, Han and Jiang  quote limits of 0.019 and 0.0028 for
$\bzti$, respectively for unpolarized and longitudinally polarized beams
with opposite-sign $e^+$ and $e^-$ polarizations. The corresponding
numbers we have from $X_1$ are 0.041 and 0.0099. The agreement is
reasonable, after
taking into account the facts that we use only one leptonic channel, and
that they employ additional experimental cuts.
The papers in \cite{biswal} also deal only with anomalous $ZZH$
couplings, and quote $3\sigma$ limits on the couplings. 
The $3\sigma$ limit they quote for
$\bzti$ is  $0.064$ for unpolarized beams, and $0.0089$ for
polarized beams. After correcting for the CL limit of $1.96\sigma$
which we use, and the inclusion of a single leptonic decay mode,
their limits are still somewhat worse. This
could be attributed to the stringent kinematic cuts imposed by them, and
to the different luminosity choice in the case of polarized beams.
Similarly, the limits quoted in \cite{biswal} for $\bzr$ and $\bzi$
are worse compared to ours by a factor of order 2 or 3
in the unpolarized as well as cases of longitudinal and transverse
polarization.

As for the case of $\gamma ZH$ couplings, comparison with earlier work
is not easy because of the different approach to parameterization of
couplings. Also, there is no work dealing in transverse polarization
with which we could make a comparison.

In the above, we have assumed a Higgs mass of 120 GeV. For larger values
of $m_H$, for larger Higgs masses, we find somewhat decreased sensitivities. 
We have also studied the sensitivities at higher c.m. energies,
possibly with a higher luminosity, and find
that in case of some observables, the sensitivity improves with
simultaneous increase in energy and luminosity.

We have not included the decay of the Higgs boson in our
analysis. For now, one could simply divide our limits by the square
root of the branching ratios and detection efficiencies. 
Including the decay will entail some loss of efficiency. 

\comment{
We have not considered scenarios with an extra neutral gauge boson $Z'$.
While it is straightforward to include a $Z'$ in or analysis, the number
of couplings would be much larger and difficult to disentangle without
studying the $s$ dependence of the cross section or asymmetries.
}

While some of these
practical questions are not addressed in this work, 
we feel that the interesting new features we found would make it
worthwhile to address them in future.


\end{document}